\documentclass{article}
\usepackage{float}
\usepackage[bitstream-charter]{mathdesign}
\usepackage{graphicx}
\usepackage[style=numeric-comp,sorting=none]{biblatex}
\begin{document}

\begin{center}{\Large \textbf{
THz Higher-Order Topological Photonics in Ge-on-Si Heterostructures\\
}}\end{center}

\begin{center}
Ian Colombo\textsuperscript{1},
Pietro Minazzi\textsuperscript{1},
Emiliano Bonera\textsuperscript{1}, 
Fabio Pezzoli\textsuperscript{1$\star$} and
Jacopo Pedrini\textsuperscript{1}
\end{center}

\begin{center}
{\bf 1} Dipartimento di Scienza dei Materiali, Università degli Studi di Milano-Bicocca and BiQuTe, via R. Cozzi 55, 20125 Milan (Italy)

$\star${fabio.pezzoli@unimib.it}

\end{center}

\section*{Abstract}
We design germanium-based higher-order topological cavities for terahertz applications by breaking the symmetry of a two-dimensional photonic crystal following the Su-Schrieffer-Heeger model. Calculations demonstrate the parity inversion of the electric field in differently deformed unit cells. The interface  between domains of opposite topology presents edge and corner modes. The former are chiral, locking light propagation to its helicity. The latter prove that Ge-based structures can be used as high-order topological photonic crystals. These findings can accelerate the development of Si-photonic components working in a spectral range of high technological interest.

\vspace{\baselineskip}

\vspace{10pt}
\noindent\rule{\textwidth}{1pt}
\tableofcontents
\noindent\rule{\textwidth}{1pt}
\vspace{10pt}

\section{Introduction}
\label{sec:intro}
The comprehension and exploitation of the topological properties of matter led to the emergence of research on topological insulators\cite{Hasan2010} and their photonic analogs, known as topological photonic crystals (TPC).\cite{Lu2014,Ozawa2019} TPCs have been shown to be promising for the fabrication of photonic integrated circuits thanks to exceptional features, e.g., directional and chiral light propagation,\cite{Lodahl2017,Wu2015,Dong2017} strong resistance to sharp bends,\cite{Zeng2020} and mathematical protection from defect-induced scattering.\cite{Wang2009} These properties are indeed expected to facilitate the implementation of advanced photonic components such as directional, polarization-dependent waveguides,\cite{Mehrabad2020waveguide,Yang2018,Parappurath2020} resonators,\cite{Mehrabad2020resonator} drop-filters\cite{Mehrabad2023} and topological lasers.\cite{Zeng2020,Han2020,Ota2018} 

Lately, higher-order topology has been gaining attention in photonics research. In contrast to conventional topological insulators, higher-order topological insulators (HOTI) present conductive states that are more than one dimension lower than the insulating state.\cite{Schindler2018,Dutt2020} 
This has led to the concept of special two-dimensional (2D) TPCs, which can feature unusual zero-dimensional (0D) corner states in addition to the conductive one-dimensional (1D)  hinge modes. The potential to exploit HOTIs to fully confine the electromagnetic field at a 0D corner and topologically protect it from undesired losses is fundamentally intriguing and strongly appealing for applications, particularly because it might drastically boost lasing emission and improve spectral purity.\cite{Han2020} 

Although crystals with a trivial photonic band structure have already found applications in the terahertz (THz),\cite{Withayachumnankul2018,Koala2022} the extension of HOTIs into such frequency range has been very limited thus far. The interest in this spectral regime comes from the inherent capacity to stream high-frequency wide-bandwidth data; \cite{Nagatsuma2016} a characteristic that offers significant prospects for the advancement of wireless communication networks beyond existing 5G standards.\cite{Yang2020,Kumar2022} In addition to telecommunications, THz waves can have far-reaching consequences in various fields, including quantum information,\cite{Yang2020,Kumar2022,Leitenstorfer2023} non-destructive imaging,\cite{Kawase2003,Jansen2010} biological sensing and diagnostics,\cite{Siegel2004,Yang2016} security and defense.\cite{Federici2005,Shen2005}
The development of efficient THz photonic components and devices is thus a compelling task where TPC and HOTIs can provide a leap forward with novel and yet untapped capabilities. 

Another crucial factor in achieving this ambitious goal is the choice of materials platform that can favor an industrial takeover while being, at the same time, suitable for the THz regime. Germanium stands out as a solution to these two problems since it offers a transparency window that is spectrally broad,\cite{Marris-Morini2018,Montesinos-Ballester2020} while being already present in microelectronic and photonic foundries. Ge-based high-quality photonic crystals (PC) can be indeed created using conventional lithography and vertical etching of thin Ge-on-Si films\cite{ElKurdi2008,Schatzl2017,Joo2021} or by exploiting self-assembly of Ge crystals directly on top of patterned Si substrates.\cite{Falub2012} This can result in high-volume production and opens the route toward monolithic integration of THz photonic components into Si chips. 

So far, literature reports have shown that Ge-on-Si heterostructures host promising, albeit non-topological, photonic properties in the near-infrared region of the electromagnetic spectrum.\cite{Pedrini2020,Pedrini2021,Falcone2022} To unfold the Ge potential in exhibiting HOTI states in the THz regime, we employ the finite elements method (FEM) to investigate photonic and topological properties including the emerge of a photonic band gap (PBG) and the topology-induced spatial confinement and directional propagation of light. In this work, we will concentrate on the model system offered by the self-assembly of micron-sized Ge-on-Si rods. Their typical in-plane arrangement can seemingly mimic 2D TPCs with a square geometry \cite{He2022,Xie2019,Chen2019,Kim2021,Han2020,Makwana2019} and their distinct optical properties\cite{Isa2016,Montalenti2018,Bergamaschini2013,Pezzoli2014,Pezzoli2016} can possibly expedite the practical realization of future, integrated HOTI devices.

\begin{figure}
    \centering
    \includegraphics[width=8cm]{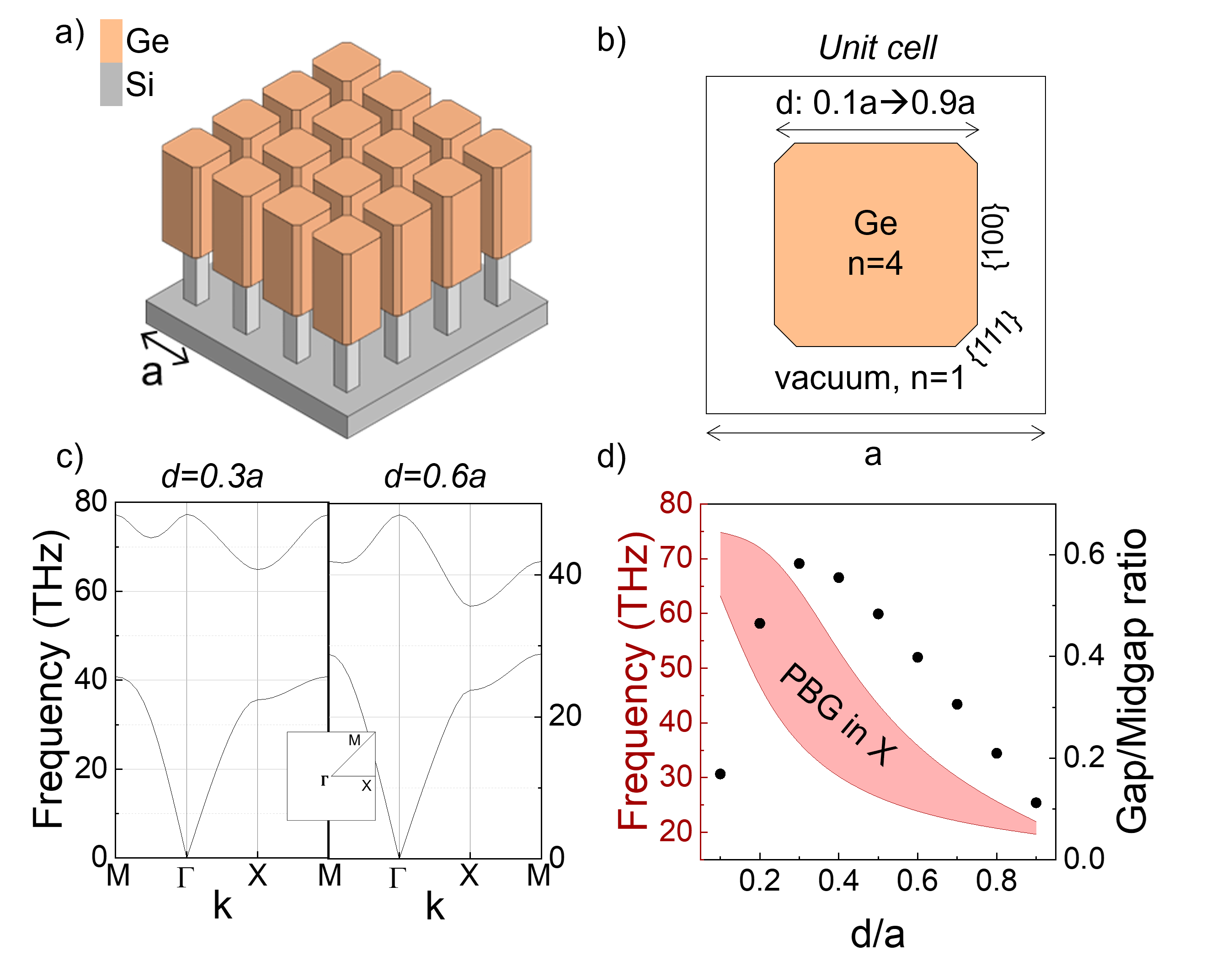}
    \caption{a) Sketch of the model photonic crystal (PC) based on a Ge (orange) on Si (grey) heterostructure (not to scale).\cite{Pedrini2021} The lattice parameter is $a$. b) Scheme of the simulated unit cell of the PC. c) Simulated bandstructure of the PC calculated using finite element method for a Ge crystal size $d=0.3a$ (left) and $d=0.6a$ (right). Inset: Irreducible Brillouin Zone of the square lattice with high simmetry points indicated. d) Size of the photonic bandgap (PBG) calculated in the $X$ point of the bandstructure (red shaded area) and gap/midgap ratio (black dots) as a function of $d$.}
    \label{Fig1}
\end{figure}

\section{Results and discussion}
\label{sec:results}

Figure \ref{Fig1}a shows the layout of a typical microstructure consisting of Ge-on-Si microcrystals. To determine the photonic bandstructure of the 2D lattice as close as possible to the experimental ones,\cite{Pedrini2021} we simulated a unit cell composed of a pseudo-octagonal Ge microcrystal, featuring both \{100\} and \{111\} facets surrounded by vacuum. The lattice parameter is $a=2$ $\mu m$ to ensure experimental feasibility with conventional fabrication processes.\cite{Pedrini2021} The size $d$ of the Ge microcrystal was varied in the FEM calculations between $0.1a$ and $0.9a$. The refractive index of Ge has been extracted from the literature\cite{Amotchkina2020} and is $n\sim4$, corresponding to the value measured in the THz region of the electromagnetic spectrum, where the extinction coefficient is zero and $n$ itself can be considered constant for the purposes of the calculations. The geometry of the unit cell, together with the structure parameters, is reported in Figure \ref{Fig1}b.

We perfomed a FEM simulation of the system eigenfrequencies with Comsol Multiphysics\cite{comsol}, using Floquet periodicity and varying the size $d$ of the microcrystal to gather information on the optimal geometric parameters of the PC. The simulation was performed for the out-of-plane electric field configuration, also known as transverse magnetic (TM) modes. The simulation sweeps the wavevector $k$ along high symmetry directions in the irreducible Brillouin Zone (IBZ), yielding the photonic bandstructure that is reported in Figure \ref{Fig1}c for two values of $d$, namely $d=0.3a$ and $d=0.6a$, corresponding to a microcrystal lateral size of 600 nm and 1200 nm, respectively. The calculated bandstructures for every value of $d$ are reported in the Supplementary Material. The bandstructures present a large PBG in the THz region of the electromagnetic spectrum.  

The bandstructures have similar shapes for different values of $d$, but its increase shifts the energy bands toward lower frequencies and apparently shrinks the amplitude of the PBG as shown in Figure \ref{Fig1}d, which reports the size of the PBG at the X high-symmetry point of the IBZ as a function of $d$. The size of the gap increases with $d$ and then decreases until it is almost negligible. This behavior is expected in 2D PCs dominated by a high refractive index material.\cite{JoannopoulosBook} To compare the size of the PBG between the different structures, we normalized the bandgap to the midgap frequency. This renormalization method allows us to compare the relative amplitude of the PBG in structures with different geometries.\cite{JoannopoulosBook} The calculation of the gap/midgap ratio in our case yields that the structure with the largest bandgap is that with $d=0.3a$. Unless otherwise noted, hereafter we refer to this specific value of $d$ (results obtained for $d=0.6a$ are nonetheless reported in the Supplementary Material). 

It should be noted that the photonic properties of the simulated system depend on the specific value of the lattice parameter $a$. However, the scaling invariance allows one to rigidly shift the energy of the PBG towards lower (higher) frequencies just by fabricating larger (smaller) unit cells. This powerful property provides great flexibility because it allows structures with a PBG in resonance with a desired frequency, e.g., the emission frequency of a quantum cascade structure. There are reports in the literature \cite{Dehlinger2000,Paul2010} showing Ge/SiGe MQWs with interband emission at $\sim 30$ THz, a value that can already be reached with the PC described in Figure \ref{Fig1}, e.g. for $d=0.8a$. The structure can be further optimized by setting $d=0.3a$, where the PBG is the largest, and increasing the lattice parameter $a$ by a factor $\sim 2$.

\begin{figure}
    \centering
    \includegraphics[width=8cm]{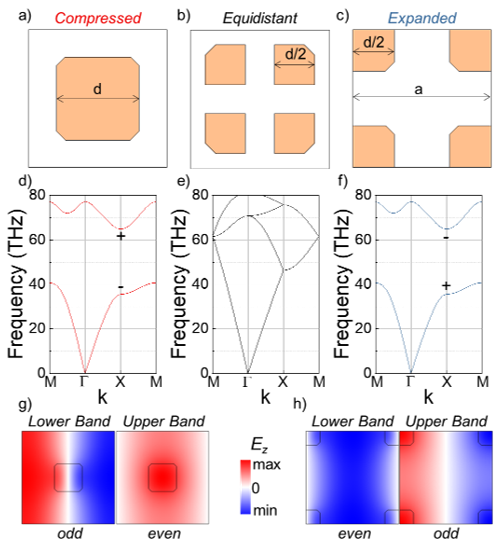}
    \caption{Scheme of the unit cell, simulated photonic bandstructure, and electromagnetic field distribution for the compressed (a,d,g), equidistant (b,e) and expanded (c,f,h) PCs when the lateral size of the Ge crystal $d$ equals 0.3 times the lattice parameter $a$. The out-of-plane component of the electromagnetic field (TM mode) is computed at the $X$ point of the IBZ. The parity of the wavefunction acts as a pseudospin, and the symmetry inversion (indicated by the $+$ and $-$) between the compressed and the expanded crystals is the fingerprint of a topological phase transition.}
    \label{Fig2}
\end{figure}

The 2D lattice composed of the semiconductor microcrystals can be seen as the periodic repetition of two different unit cells. The two structures can be considered the extreme case of a photonic extension of a 2D Su-Schrieffer-Heeger (SSH) lattice,\cite{Su1979,Liu2017,Xie2019} where a unit cell composed of four elements equidistant from both the center and the vertex of the cell is distorted, as shown in Figure \ref{Fig2}. The first unit cell has a microcrystal with lateral size $d$ at the center of the cell, as shown in Figure \ref{Fig1}b or Figure \ref{Fig2}a, and will from now on be referred to as \textit{compressed}. The other structure consists of four quarters of a microcrystal with a width $\frac{d}{2}$ placed at the corners of the cell, as shown in Figure \ref{Fig2}c. We will refer to this structure as \textit{expanded}. The \textit{equidistant} unit cell structure is reported in Figure \ref{Fig2}b.
\\
The bandstructures of the described lattices are reported in Figure \ref{Fig2}d-f. The one of the \textit{equidistant} PC (reported in Figure \ref{Fig2}e) is gapless and shows a pseudo-Dirac point at the M and X high-symmetry points. The deformation of the unit cell opens a gap, as expected in the SSH model, and yields two identical photonic bandstructures for the \textit{compressed} and \textit{expanded} PCs. It is important to highlight that in a SSH model the band dispersion does not change with the inversion of the intra- and inter-cellular distances between the elements composing the unit cell, but the symmetry of the eigenfunctions is different, as they possess opposite parity.\cite{Liu2017,Xie2019} To gather further insights on the bandstructure of the \textit{expanded} and \textit{compressed} PCs, we calculated the out-of-plane electric field distribution $E_z$ (TM mode) for such unit cells. Particularly, we investigate the $E_z$ distribution at the X point of the bandstructure, where the PBG opens up. The $E_z$ distribution maps are reported in Figure \ref{Fig2}g,i. Here, the \textit{compressed} PC presents an even $E_z$ distribution in the lower band and an odd distribution in the high-energy band. The opposite occurs in the \textit{expanded} structure. This parity inversion confirms the equivalence of the two PC structures to a 2D SSH model. Therefore, the \textit{compressed} and \textit{expanded} PC belongs to distinct topological phases, where the parity of the bands can be considered as the topological invariant. In particular, the \textit{compressed} structure is an ordinary insulator, while the \textit{expanded} is topologically nontrivial.
 
\begin{figure*}
    \centering
    \includegraphics[width=14cm]{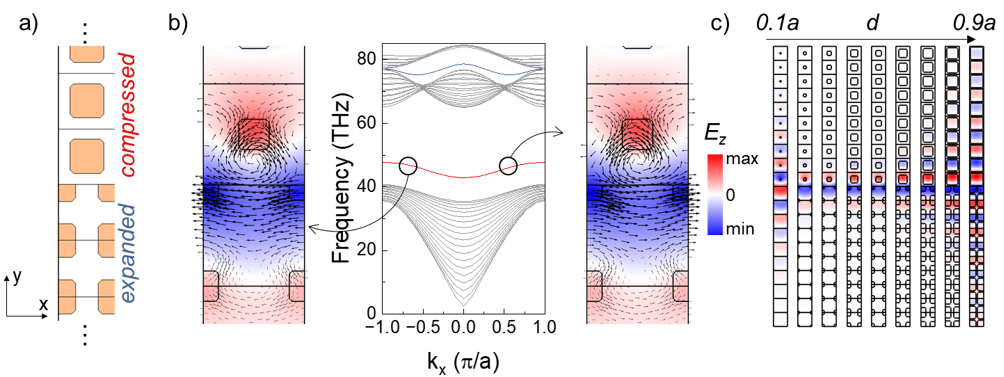}
    \caption{a) Schematics of a supercell consisting of a line interface between a compressed and expanded PCs. b) Calculated bandstructure of the supercell along the $x$ direction. The bandstructure presents bulk bands (grey) with two sizeable gaps in which localized modes are present (red and blue curves). The modes are confined at the interface of the two regions of the PCs. The arrows overlaid on the electromagnetic field distribution underline the directionality of the propagation. c) Spatial distribution of the out-of-plane component of the electromagnetic field ($E_z$) in the supercell as a function of the lateral size of Ge $d$. The supercells are stacked horizontally as $d$ increases from $0.1a$ to $0.9a$, where $a$ is the lattice parameter.}
    \label{Fig3}
\end{figure*}

The clearest evidence of the presence of a topological transition is the emergence of spatially confined guided modes at the boundary between two domains with different band topology.\cite{Wu2015,Khanikaev2017,Zeng2020,Gong2020,Shalaev2019} Figure \ref{Fig3}a reports the schematic of an interface between the two PCs characterized by distinct topological invariants. For its characterization we designed a so-called \textit{supercell} composed of a ribbon of 20 unit cells where the top (bottom) 10 unit cells are \textit{compressed} (\textit{expanded}). In other words, the top half of the supercell is an ordinary insulator, while the bottom half is topologically nontrivial. The FEM simulation of this structure is performed with periodic conditions along the $x$ direction, and the eigenfrequencies are calculated as a function of $k_x$, from $-\frac{\pi}{a}$ to $\frac{\pi}{a}$. A perfectly matched layer is used as the boundary condition for the top and bottom of the ribbon to simulate an infinite PC. The resulting bandstructure is shown in Figure \ref{Fig3}b. It presents a large number of bulk modes and two energy gaps, the larger of which covers the interval between 41 and 65 THz, while a second, non complete one is at around 75 THz. For the scope of this work, we focus
on the full PBG at lower energy. The bandgap frequencies are the same as those calculated for the bulk unit cells along the $\Gamma-X$ direction (see Figure \ref{Fig2}). The presence of a single mode in the PBG, located at $\sim45$ THz, is a fingerprint of the interface of two phases with a different topological invariant. Such a mode is spatially localized at the interface of the two domains, as is shown by the plot of $E_z$ (see Figure \ref{Fig3}b), with the electric field mostly penetrating the high-index structure. The arrows overlaid on the $E_z$ map are the local Poynting vectors that represents the direction of propagation of the electromagnetic wave. The representation of the Poynting vector allows us to underline the presence of unidirectional propagating modes, that can be selectively coupled through helical excitation.\cite{Ozawa2019,Wu2015,Khanikaev2017} 
Figure \ref{Fig3}c shows that when $d$ is varied the imbalance between the air and Ge fractions affects the confinement of the edge mode, so that the field is almost perfectly localized within the two interfacial unit cells only for $d$ ranging from $0.2a$ to $0.5a$. 

\begin{figure*}
    \centering
    \includegraphics[width=14cm]{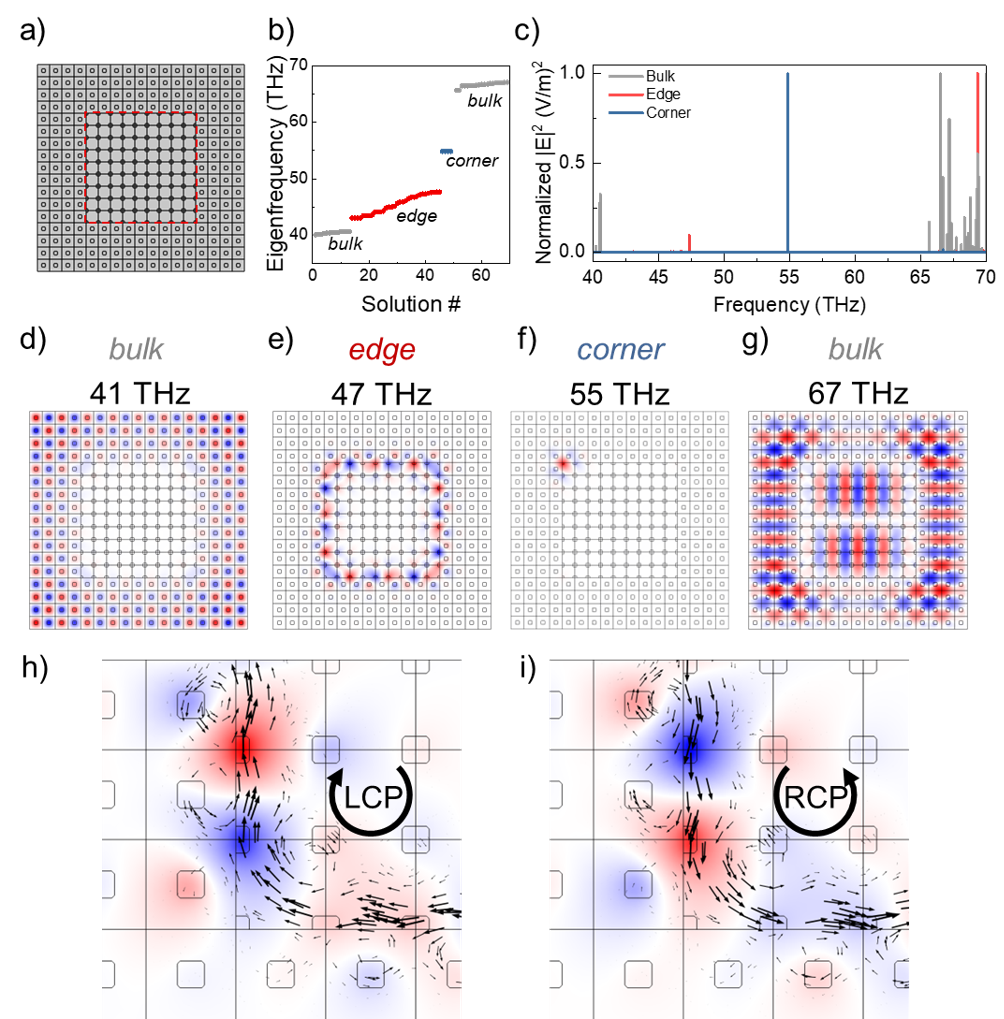}
    \caption{a) Schematics of a resonator composed of a square interface between an expanded PC surrounded by a compressed PC ($d=0.3a$). The interface is marked with a red dashed line. b) Eigenfrequency values of the resonator as a function of the solution number. Four groups can be identified that correspond to bulk modes (low- and high-energy, grey), edge (red), and corner (blue) modes. c) Normalized field intensity as a function of the frequency, highlighting the bulk, edge and corner modes. d-g) Distribution of the out-of-plane component of the electric field at four significant frequencies corresponding to a low-energy bulk mode (d), edge mode (e), corner mode (f), and a high-energy bulk mode (g). h,i) Electromagnetic field $E_z$ distribution at the bottom left corner of the resonator, when the resonator is excited with left (h) and right (i) circularly polarized light. The arrows at the interface between the topologically distinct regions are the Poynting vectors, highlighting a direct correspondence between light polarization and the direction of propagation.}
    \label{Fig4}
\end{figure*}

The demonstration of the presence of optical modes at the interface between domains suggests a possible application of Ge-on-Si photonic architectures as on-chip THz waveguides in topological circuits. We can further extend our results by designing a 2D device that could also exploit the generation of higher-order topological modes at the intersection between such hinge modes. Figure \ref{Fig4}a introduces a resonator composed of a square of the \textit{expanded} PC having a side of 9-unit cells, surrounded by a cladding frame consisting of 4-unit cells of the \textit{compressed} PC defining an interface that supports the mode described in Figure \ref{Fig3}.
The solutions of the eigenvalue analysis for the resonator are separated in four well-defined frequency regions, as shown in Figure \ref{Fig4}b,c. The nature of these modes can be determined by analyzing the electric field distribution, as shown in Figure \ref{Fig4}d-g. The electromagnetic field maps for solutions for frequencies $<41$ THz (see Figure \ref{Fig4}d) and $>65$ THz (see Figure \ref{Fig4}g) clearly demonstrate the bulk nature of the modes, that permeate vast regions of the PC. 
In the frequency range pertaining to the PBG two well separated sets of solutions are present at $\sim 47$ THz and at $55$ THz. First, we focus on the four degenerate modes at $55$ THz that dominate the energy density spectrum reported in Figure \ref{Fig4}c. The map of the electric field distribution, reported in Figure \ref{Fig4}f, shows that these are extremely localized 0D corner modes. Their existance demonstrates that the structure described in this work is a higher-order TPC characterized by a bulk-edge-corner correspondence.\cite{Xie2018} Moreover, localized corner modes are extremely interesting for their strong confinement properties and can be exploited for their possible applications to devices that need high-quality factor resonators such as light emitters, sensors, and non-linear systems.\cite{Chen2019,Xie2019,Zhang2020,Bravo-Abad2007}

We now focus on the lower energy modes, found at frequency around $47$ THz. The electromagnetic field distribution shows that these are edge modes confined at the interface between the trivial and topological PC structures. Their study can give further insight on the topological properties of the PC and how they influence the propagation of light at the interface between the two topologically-distinct domains.
As described above, a characteristic property of TPCs is the directional propagation of light, which is related to its degree of circular polarization. To demonstrate this feature, we simulated the propagation of circularly polarized light by using an array of opportunely spaced phased dipoles localized at the interface between the topologically distinct regions.\cite{Gao2018} The overlay of the Poynting vector on the electromagnetic field map, shown in Figure \ref{Fig4}h-i, demonstrates how the propagation is strongly directional and locked to the degree of circular polarization, allowing chiral propagation at the interface of the PCs in the THz range.

\section{Conclusions}
\label{sec:conclusions}
We demonstrated the possibility of achieving higher-order topological effects in the THz regime in a PC composed of group IV heteroepitaxial microstructures. Such a HOTIs can be utilized for the development of elemental components of photonic circuitries such as resonators and waveguides. By combining Ge-based heterostructures with the intrinsic scalability of PCs one can obtain devices working in a wide range of frequencies, possibly from mid-infrared to the THz. Furthermore, the capacity to embed THz emitters in the microstructures in the form of Ge/SiGe quantum  wells might open a pathway to realize integrated, topological lasers with a small footprint and high throughput that operate within technologically  relevant spectral regions.

\section*{Acknowledgements}
The authors thank A. Marzegalli for technical assistance and L. Miglio for fruitful discussions.

\paragraph{Funding information}
This work has been funded by the European Union's Horizon Europe Research and Innovation Programme under agreement 101070700. Support from PNRR MUR project PE0000023-NQSTI is also acknowledged. J.P. acknowledges financial support from FSE REACT-EU (grant 2021-RTDAPON-144). 

\printbibliography
\pagebreak

\begin{appendix}
{\large\section*{Supplementary Material}}

In this supplementary material are reported the bandstructure of the PC as a function of the Ge microcrystal size $d$, the demonstration of topological inversion for $d=0.6a$, the appearance of the localized edge mode with chiral conductivity for $d=0.6a$, and the design of the square resonator with the eigenfrequency analysis and electromagnetic field distribution for $d=0.6a$.

\begin{figure*}[h]
    \centering
    \includegraphics[width=12cm]{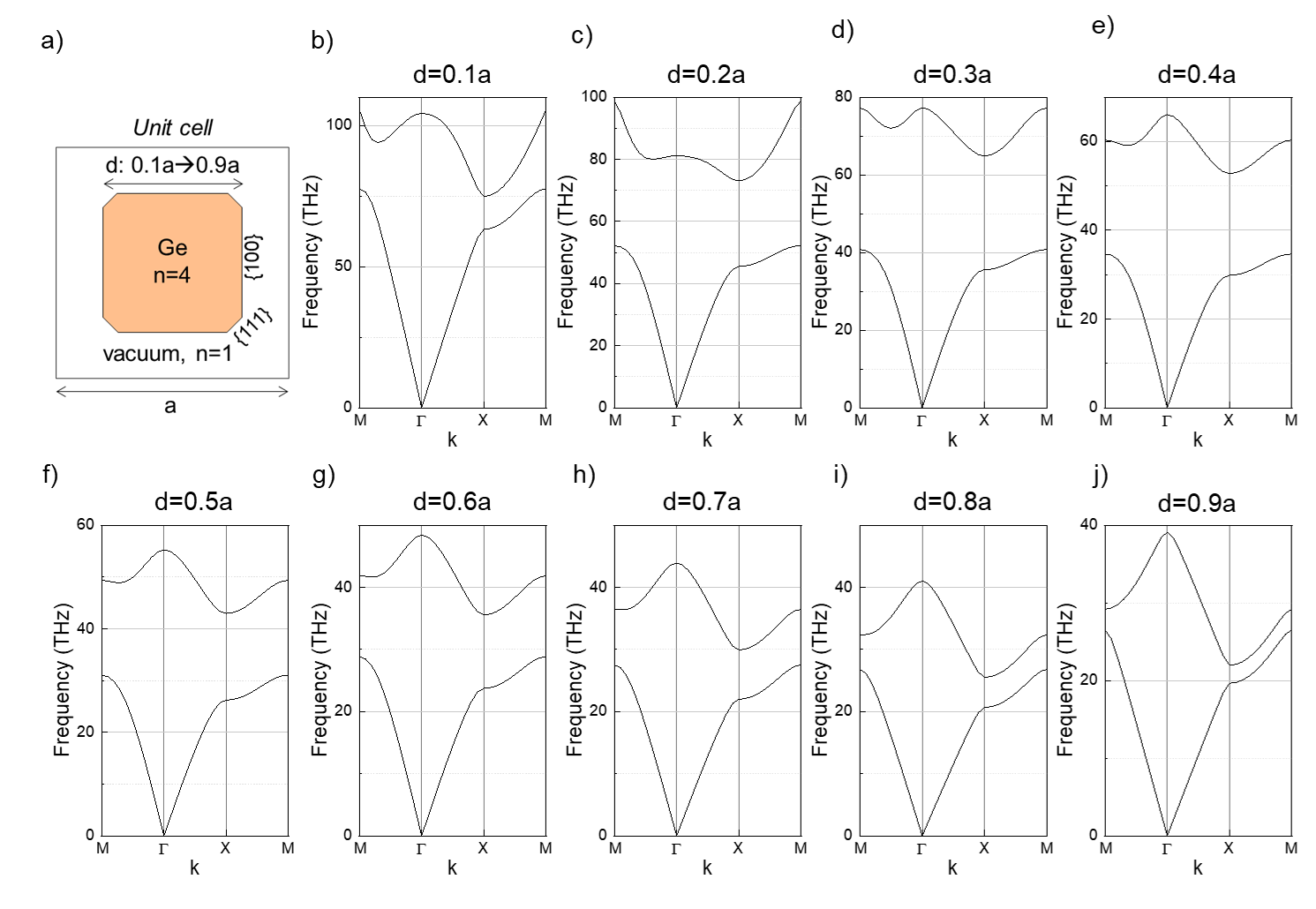}
    \caption{photonic bandstructure calculated with the finite element method for the PC described in a) as a function of the ratio between the size of the Ge element d and the lattice parameter a.}
    \label{FigS1}
\end{figure*}

\begin{figure*}[h]
    \centering
    \includegraphics[width=12cm]{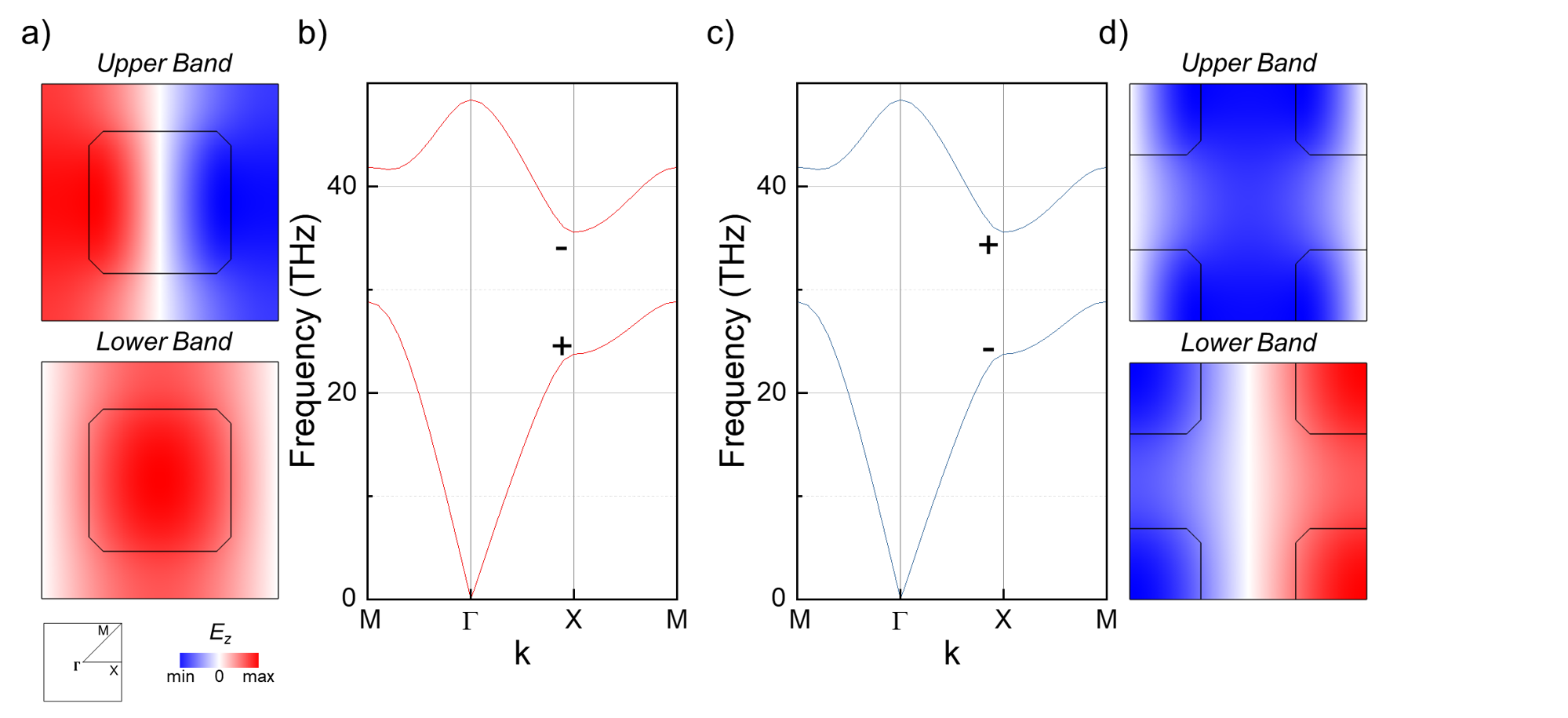}
    \caption{electromagnetic field distribution and simulated photonic bandstructure for the compressed (a,b) and expanded (c,d) PCs when the lateral size of the Ge crystal d equals 0.6 times the lattice parameter a. The out-of-plane component of the electromagnetic field (TM mode) is computed at the X point of the IBZ. The parity of the wavefunction acts as a pseudospin, and the symmetry inversion (indicated by the + and -) between the compressed and the expanded crystals is the fingerprint of a topological phase transition.}
    \label{FigS2}
\end{figure*}

\begin{figure*}[h]
    \centering
    \includegraphics[width=12cm]{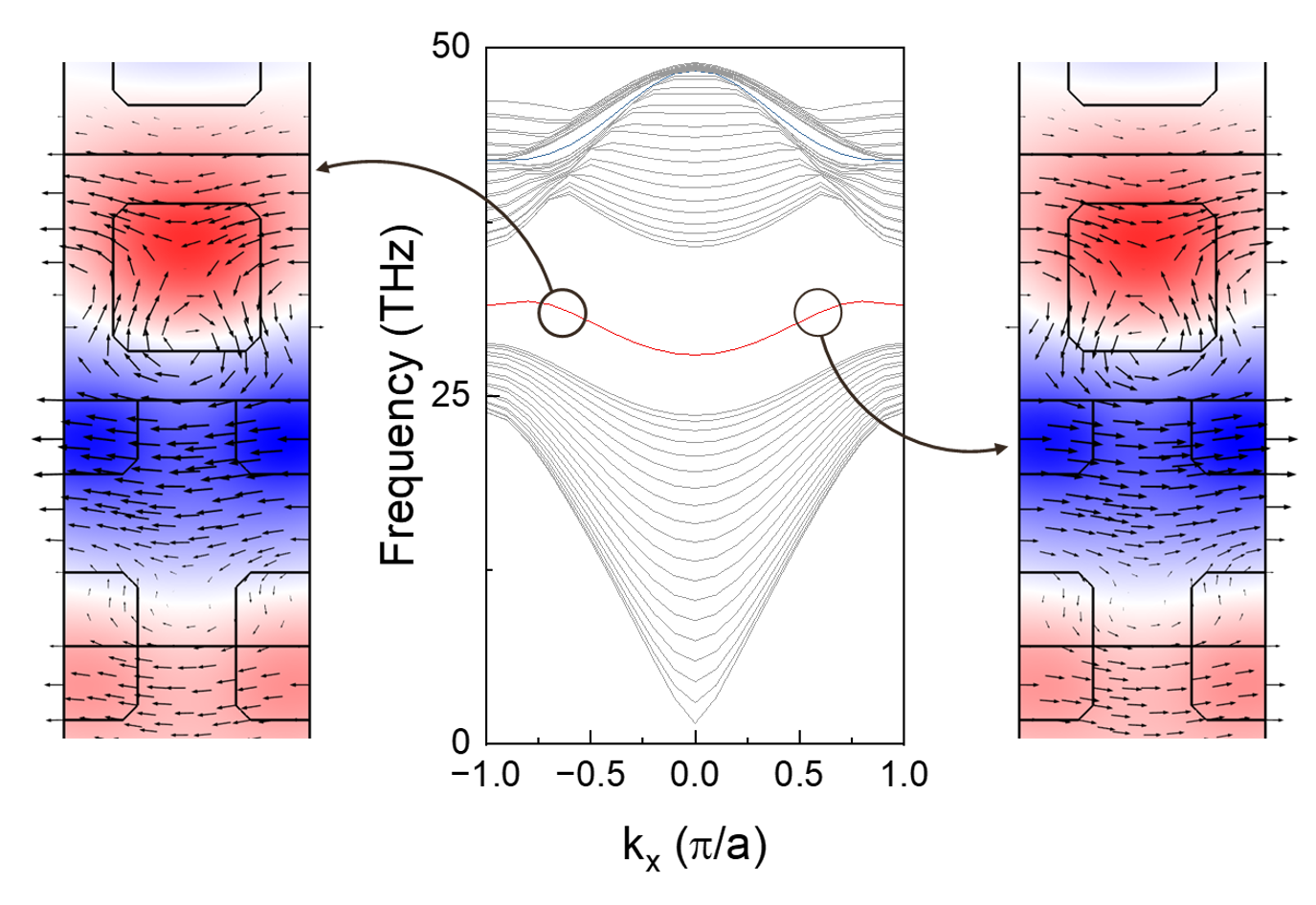}
    \caption{Calculated bandstructure of a supercell, composed by a trivial domain (top) interfaced with a topological domain (bottom), along the x direction. The bandstructure presents bulk bands (grey) with two sizeable gaps in which localized modes are present (red and blue curves). The modes are confined at the interface of the two regions of the PCs. The arrows overlaid on the electromagnetic field distribution underline the directionality of the propagation.}
    \label{FigS3}
\end{figure*}

\begin{figure*}[h]
    \centering
    \includegraphics[width=12cm]{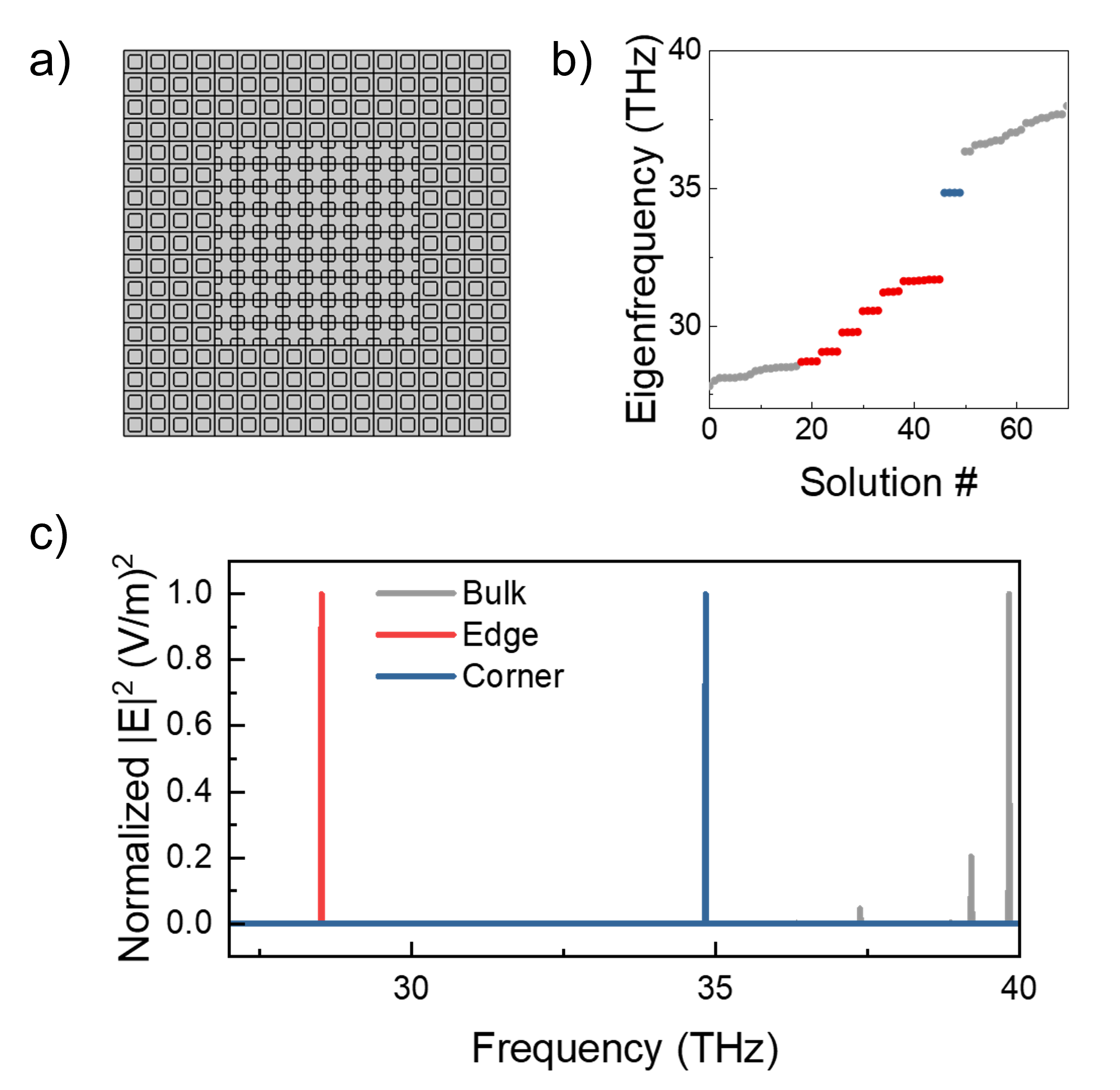}
    \caption{a) Schematics of a resonator composed of a square interface between an expanded PC surrounded by a compressed PC (d = 0.6a). b) Eigenfrequency values of the resonator as a function of the solution number. Four groups can be identified that correspond to bulk modes (low- and high-energy, grey), edge (red), and corner (blue) modes. c) Normalized field intensity as a function of the frequency, highlighting the bulk, edge and corner modes.}
    \label{FigS4}
\end{figure*}

\begin{figure*}[h]
    \centering
    \includegraphics[width=12cm]{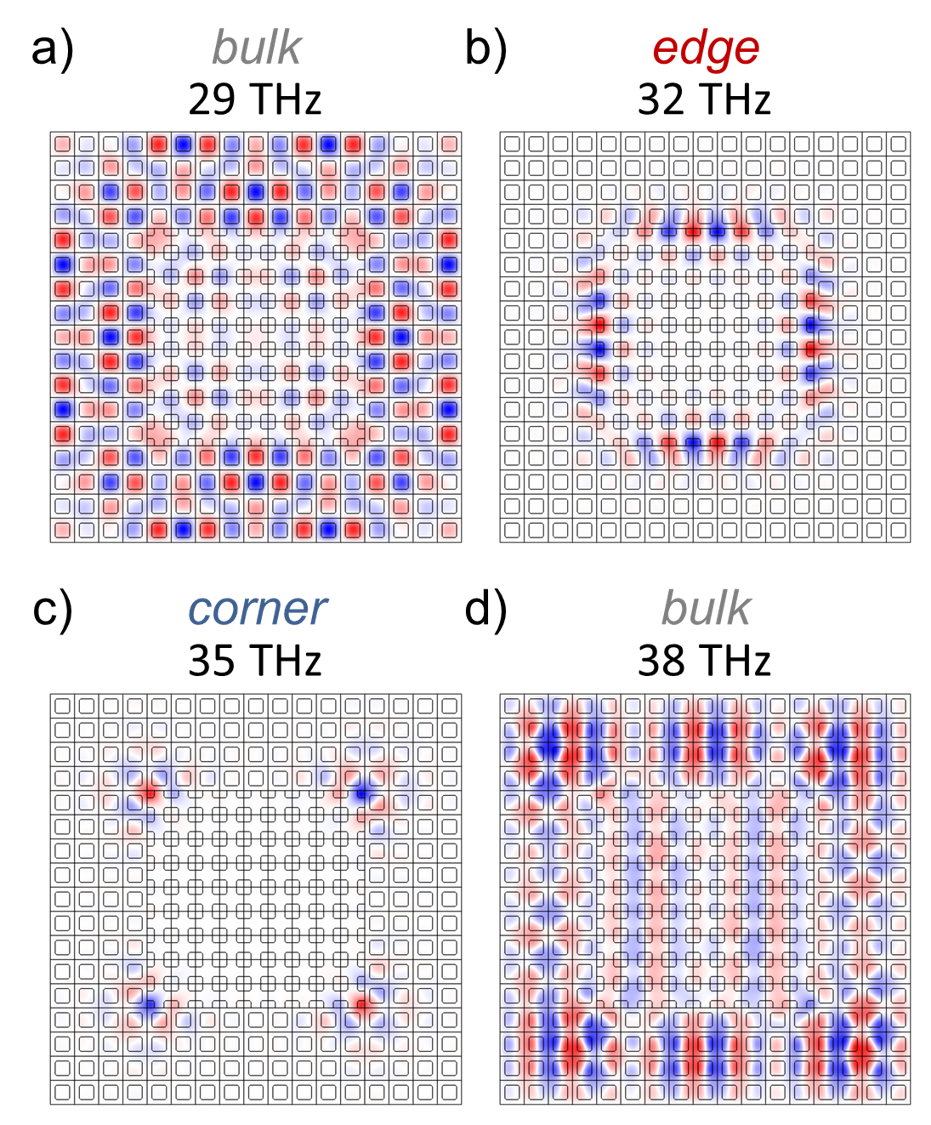}
    \caption{Distribution of the out-of-plane component of the electric field at four significant frequencies corresponding to a low-energy bulk mode (a), edge mode (b), corner mode (c), and a high-energy bulk mode (d).}
    \label{FigS5}
\end{figure*}
\end{appendix}

\end{document}